\documentclass[jcp,aip,twocolumn,reprint,longbibliography,floatfix,superscriptaddress]{revtex4-1}
\usepackage{amsmath,txfonts,graphicx,upgreek,xfrac}
\usepackage[bookmarks=true,colorlinks=true,urlcolor=blue,linkcolor=blue,citecolor=blue]{hyperref} 
\usepackage{multirow}
\newcommand{\FT}{\mathcal{F}}

\newcommand{\hamKS}{\mathcal{H}_{\mathrm{KS}}}
\newcommand{\kick}{\mathcal{K}}
\newcommand{\Proj}{\mathcal{T}}

\newcommand{\rv}{\textbf{r}}
\newcommand{\BigO}{\mathcal{O}}
\newcommand{\eq}{\hat{\textbf{e}}_{\textbf{q}}}
\newcommand{\Deltax}{\Delta_{x}}
\usepackage{xfrac}
\renewcommand{\Im}{\operatorname{Im}}

\newcommand{\LCAOTDDFTkomega}{LCAO-TDDFT-$k$-$\omega$}
\newcommand{\PWTDDFTkomega}{PW-TDDFT-$k$-$\omega$}
\newcommand{\LCAOTDDFTromega}{LCAO-TDDFT-$r$-$\omega$}
\newcommand{\LCAOTDDFTrt}{LCAO-TDDFT-$r$-$t$}
\newcommand{\RSTDDFTrt}{RS-TDDFT-$r$-$t$}

\begin{document}

\title{Optical Excitations of Chlorophyll~\emph{a} and \emph{b} Monomers and Dimers}

\author{Mar\'{\i}a Rosa Preciado-Rivas}
\affiliation{ School of Physical Sciences and Nanotechnology, Yachay Tech University, Urcuqu\'{\i} 100119, Ecuador}
\author{Duncan John Mowbray}
\email{duncan.mowbray@gmail.com}
\affiliation{ School of Physical Sciences and Nanotechnology, Yachay Tech University, Urcuqu\'{\i} 100119, Ecuador}
\affiliation{Department of Applied Mathematics, University of Waterloo, Canada N2L 3G1}
\affiliation{Nano-Bio Spectroscopy Group and ETSF Scientific Development Centre, Departamento de F\'{\i}sica de Materiales, Universidad del Pa\'{\i}s Vasco UPV/EHU, E-20018 San Sebasti\'{a}n, Spain}
\author{Keenan Lyon}
\affiliation{Department of Applied Mathematics, University of Waterloo, Canada N2L 3G1}
\author{Ask Hjorth Larsen}
\affiliation{Nano-Bio Spectroscopy Group and ETSF Scientific Development Centre, Departamento de F\'{\i}sica de Materiales, Universidad del Pa\'{\i}s Vasco UPV/EHU, E-20018 San Sebasti\'{a}n, Spain}
\author{Bruce Forbes Milne}
\affiliation{CFisUC, Department of Physics, University of Coimbra, Rua Larga, 3004-516 Coimbra, Portugal}
\affiliation{Nano-Bio Spectroscopy Group and ETSF Scientific Development Centre, Departamento de F\'{\i}sica de Materiales, Universidad del Pa\'{\i}s Vasco UPV/EHU, E-20018 San Sebasti\'{a}n, Spain}
\affiliation{Coimbra Chemistry Center, Department of Chemistry, University of Coimbra, Rua Larga, 3004-535 Coimbra, Portugal}

\begin{abstract}
  A necessary first step in the development of technologies such as artificial photosynthesis is understanding the photoexcitation process within the basic building blocks of naturally-occurring light harvesting complexes (LHCs).  The most important of these building blocks in biological LHCs such as LHC II from green plants are the chlorophyll~\emph{a} (Chl~\emph{a}) and chlorophyll~\emph{b} (Chl~\emph{b}) chromophores dispersed throughout the protein matrix.  However, efforts to describe such systems are still hampered by the lack of computationally efficient and accurate methods that are able to describe optical absorption in large biomolecules.    In this work we employ a highly efficient linear combination of atomic orbitals (LCAOs) to represent the Kohn--Sham (KS) wave functions at the density functional theory (DFT) level and perform time dependent density functional theory (TDDFT) in either the reciprocal space and  frequency domain (\LCAOTDDFTkomega{}) or real space and time domain (\LCAOTDDFTrt{}) calculations of the optical absorption spectra of Chl~\emph{a} and \emph{b} monomers and dimers.  We find our \LCAOTDDFTkomega{} and \LCAOTDDFTrt{} calculations reproduce results obtained with a plane wave (PW) representation of the KS wave functions (\PWTDDFTkomega{}), but with a significant reduction in computational effort.  Moreover, by applying the GLLB-SC derivative discontinuity correction $\Deltax$ to the KS eigenenergies, with both \LCAOTDDFTkomega{} and \LCAOTDDFTrt{} methods we are able to semi-quantitatively reproduce the experimentally measured photoinduced dissociation (PID) results. This work opens the path to first principles calculations of optical excitations in macromolecular systems.
\end{abstract}
\maketitle

\section{Introduction}

Chlorophyll~\emph{a} (Chl~\emph{a} or C$_{55}$H$_{72}$MgN$_4$O$_5$) and chlorophyll~\emph{b} (Chl~\emph{b} or C$_{55}$H$_{70}$MgN$_4$O$_6$) \cite{Moss88,Chlorophylls}  are the fundamental functional units of the light harvesting complex (LHC II) \cite{ChromophoreSpectrum} present in green plants.  For this reason, understanding the photoexcitation process within Chl~\emph{a} and \emph{b} is of great importance in the development of technologies such as those involved in the optimization of food crop production \cite{Long2014,Ray2013} and conversion of solar radiation into a usable form of energy directly through methods such as conventional solar cells \cite{Ruban2011,Fleming2012} or via subsidiary technologies such as photosynthetically-driven (bio)reactor systems for hydrogen combustion \cite{Allakhverdiev2010,Krassen2009,Berardi2014,Searle2014}.  Moreover, such information is more generally applicable to the \emph{in silico} design and optimization of dye-sensitized solar cells \cite{RevDSSCs}, organic photovoltaic cells \cite{RevOPVs}, photocatalytic systems \cite{RevPhotoCatalSys}, optoelectronic devices \cite{RevOptoElecDev}, and plasmonics \cite{RevPlasmonics}.

Although much progress has been recently made in both the experimental measurement of individual monomer and dimer Chl~\emph{a} and \emph{b} spectra \cite{Bruce1,Bruce2,Chla2}, and their theoretical description at the time-dependent density functional theory (TDDFT) \cite{RungeGross} level \cite{Hedayatifar2016,Linnanto2006,Dreuw2004,Parusel2000}, the lack of reasonably accurate yet highly efficient computational methods has hampered efforts to describe the optical absorption of large Chl-containing biomolecules. Initial attempts to investigate the optical absorption characteristics of biomacromolecules such as the LHC II using first-principles electronic structure methods have helped to clarify several aspects of the functioning of these systems\cite{Joaquim,Konig2011}. However, these calculations neglected the protein matrix, and thus failed to fully explain the role played by the surrounding proteins in photosynthesis and why they are produced.  Moreover, the computational resources required for a complete treatment of systems of this size lie considerably beyond what is generally available to most researchers.  

On the one hand, methods based on the Kohn--Sham (KS) density of states \cite{KohnSham}, while being quite efficient, often underestimate energy gaps by more than half.  This is because an independent-particle picture fails to describe electronic screening of the excited states \cite{OurJACS}.  On the other hand, quasiparticle-based calculations of spectra from the Bethe--Salpeter equation (BSE) \cite{BetheSalpeterEqn}, while often achieving quantitative accuracy, are extremely heavy computationally \cite{AngelGWReview}.  As a result, only recently have even the smallest dye-sensitized solar cells (DSSC) been described at the BSE level \cite{Catechol}.  Moreover, such methods are intrinsically ill-suited to the description of isolated and/or non-periodic systems.   

Although TDDFT \cite{RungeGross,TDDFTBook} real time \cite{TDDFTrt,TDDFTrt2,JCTCRT-TDDFTBig2017,JCTCRT-TDDFTEmbed2017} and frequency domain \cite{Casida1995,TDDFTRevCasida2009,response1,response2,DuncanGrapheneTDDFTRPA} calculations provide an attractive alternative, implementations based on real-space (RS) or plane-wave (PW) representations of the KS orbitals \cite{KohnSham} are both computationally expensive, and exhibit a strong exchange and correlation (xc) functional dependence for their accuracy.  Time propagation RS (\RSTDDFTrt) calculations require time steps much shorter than what is needed to resolve the features of the spectra in order to ensure the stability of the calculation \cite{Joaquim}.  Such instabilities of RS calculations may be related to their freedom in representing the KS wavefunctions, which are only constrained by the grid spacing.

In this work we employ linear combinations of atomic orbitals (LCAOs) to provide a more efficient representation of the KS orbitals, while retaining the accuracy of PW-based TDDFT reciprocal space and frequency domain \PWTDDFTkomega{} calculations of the optical absorption.   Moreover, we use the recently developed derivative discontinuity correction based on the exchange part of the GLLB-SC \cite{GLLBSC} functional $\Deltax$ to correct the KS eigenenergies and provide a semi-quantitative agreement with experimental measurements \cite{Castelli2012GLLBsc2,PreciadoSWCNTs,LCAOTDDFTKeenan}.  Furthermore, the constraints imposed by an LCAO representation of the KS wavefunctions may be expected to improve the stability of time-propagation TDDFT calculations (\LCAOTDDFTrt), allowing one to use larger time steps.  However, the reliability of LCAO-TDDFT is inherently basis set dependent \cite{LCAOBasisSets}.  This makes an assessment and benchmarking for the fundamental functional units with \PWTDDFTkomega{} calculations and photoinduced dissociation (PID) experiments essential before applying \LCAOTDDFTkomega{} or \LCAOTDDFTrt{} to the complete macromolecular system.

By applying LCAO-TDDFT methods \cite{GlanzmannTDDFTRPA,PreciadoSWCNTs,LCAOTDDFTKeenan,LCAOTDDFT} to the Chl~\emph{a} and \emph{b} monomer and dimer systems, we may clearly explain their advantages and disadvantages for describing light-harvesting systems, with the aim of applying these methods to macromolecules such as the complete LHC II.  Note that since experimental measurements of the Chl~\emph{b} dimer are currently unavailable, we have restricted consideration to the Chl~\emph{a} dimer herein.

This paper is organized as follows.  In Sec.~\ref{Sect:Methodology} we describe the computational parameters employed to model the Chl~\emph{a} and \emph{b} systems and provide a theoretical background to \LCAOTDDFTkomega{} \cite{GlanzmannTDDFTRPA,PreciadoSWCNTs,LCAOTDDFTKeenan} and \LCAOTDDFTrt{} \cite{LCAOTDDFT} calculations in the projector augmented wave (PAW) method.  In Sec.~\ref{Sect:ResultsDiscussion}, we perform a basis set convergence test and direct comparison with \PWTDDFTkomega{} calculations and PID measurements for our \LCAOTDDFTkomega{} calculations of Chl~\emph{a} and \emph{b}, model the excitonic density of their first four bright excitations $\omega_i$ using the electron hole density difference $\Delta \rho(\rv,\omega_i)$ from \LCAOTDDFTkomega{}, compare our \LCAOTDDFTrt{} calculations with PID measurements for Chl~\emph{a} and \emph{b}, compare both \LCAOTDDFTkomega{} and \LCAOTDDFTrt{} calculations with PID measurements for the Chl~\emph{a} dimer (Chl~\emph{a})$_2$, and finally tabulate and compare our results for the Q band with those available in the literature \cite{Bruce1,Chla2}.  We then provide concluding remarks in Sec.~\ref{Sect:Conclusions}.  Atomic units ($\hslash = e = m_e = a_0 = 1$) are employed throughout unless stated otherwise.  

\section{METHODOLOGY}\label{Sect:Methodology}

\begin{figure}[!t]
  \includegraphics[width=\columnwidth]{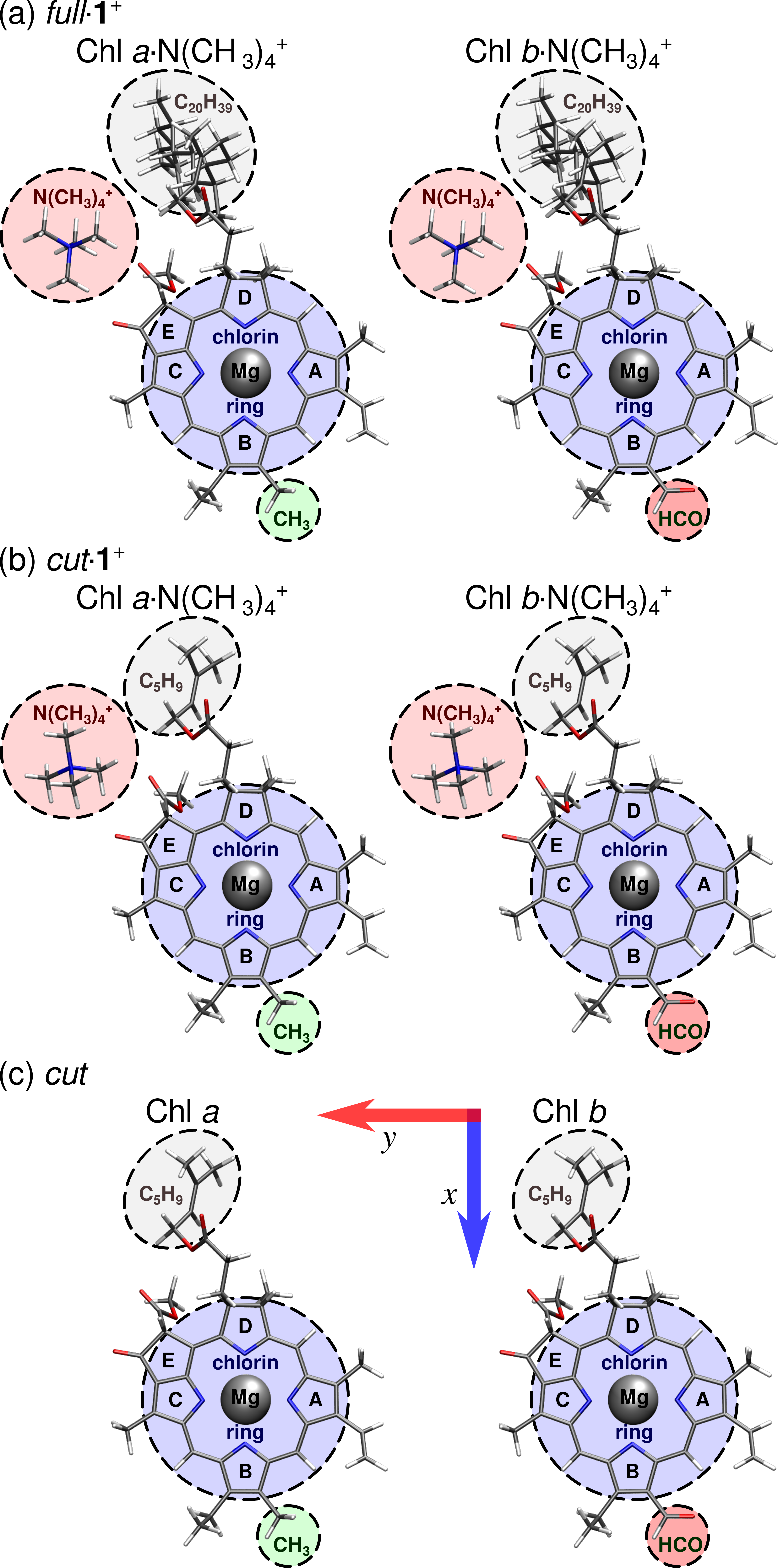}
  \caption{Schematics of (a) \emph{full}$\cdot$\textbf{1}$^+$, (b) \emph{cut}$\cdot$\textbf{1}$^+$, and (c) \emph{cut}  chlorophyll~\emph{a} (Chl~\emph{a}) and chlorophyll~\emph{b} (Chl~\emph{b}) structures with Mg atom, chlorin ring (blue), methyl group (CH$_3$, green) or aldehyde (HCO, red) groups, full (C$_{20}$H$_{39}$, grey) or cut (C$_{5}$H$_{9}$, grey) hydrocarbon chain, and a tetramethylammonium charge tag (N(CH$_3$)$_4^+$, red).   Labelling of the rings (\textbf{A}--\textbf{E}) and orientation of the $x$ and $y$ polarization axes (blue and red arrows) is according to IUPAC-IUB nomenclature \cite{Chlorophylls,Moss88}.   Mg, C, O, N, and H atoms are depicted in silver, grey, red, blue, and white. Structures in (a) \emph{full}$\cdot$\textbf{1}$^+$ are based on those provided in Ref.~\citenum{Bruce1}.}\label{MonomerStructure}
\end{figure}

\begin{figure}
  \includegraphics[width=0.4\columnwidth]{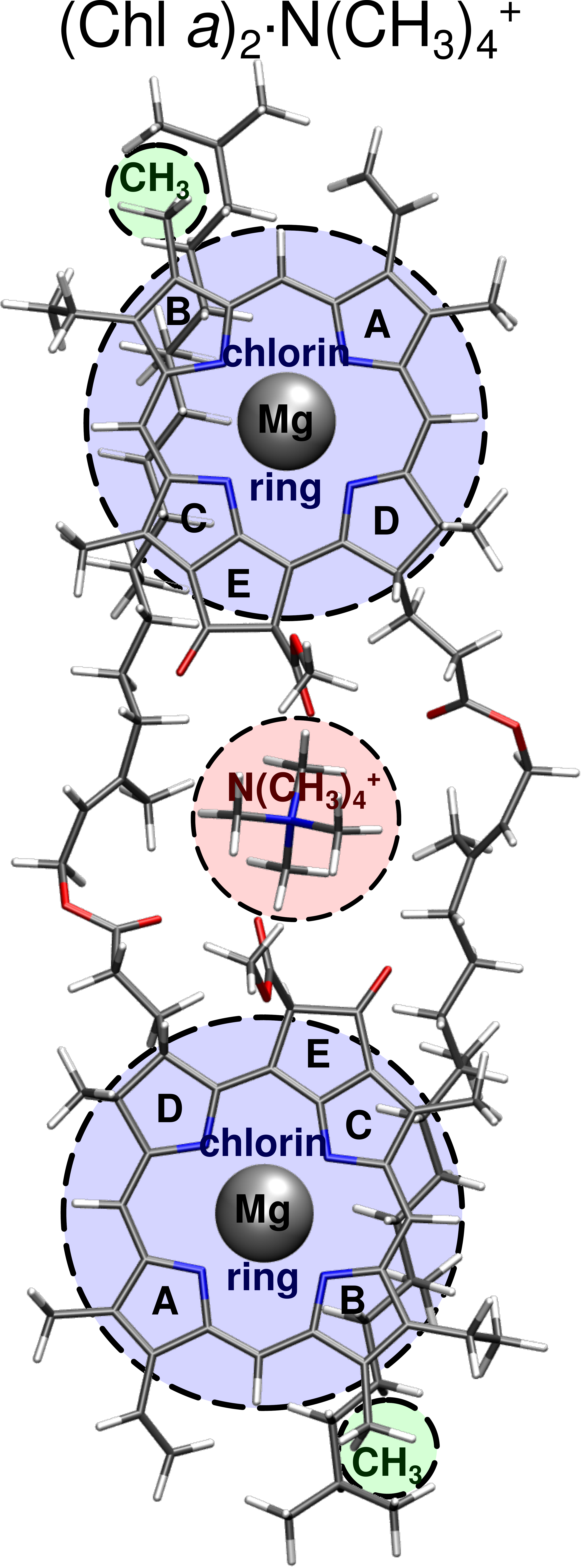}
  \caption{Schematic of the chlorophyll~\emph{a} dimer (Chl~\emph{a})$_2$ structure with Mg atom, chlorin ring (blue), methyl group (CH$_3$, green), and a tetramethylammonium charge tag (N(CH$_3$)$_4^+$, red).  Labelling of the rings (\textbf{A}--\textbf{E}) is according to IUPAC-IUB nomenclature \cite{Chlorophylls,Moss88}. Mg, C, O, N, and H atoms are depicted in silver, grey, red, blue, and white. The structure is based on the linear configuration provided in Ref.~\citenum{Chla2}.}\label{DimerStructure}
\end{figure}

All DFT \cite{HohenbergKohn,KohnSham} calculations were performed using the PAW \cite{PAW,PAW2,CarstenThesis} implementation within the GPAW code \cite{GPAW,GPAWRev,GPAWLCAO}.  We employed the generalized gradient approximation (PBE \cite{PBE}) for the exchange and correlation functional. To represent the electron density and wave functions we employed either LCAOs \cite{GPAWLCAO} with a double-$\zeta$-polarized (DZP) basis set, after performing convergence tests with basis sets of varying quality up to a quadruple-$\zeta$-polarized (QZP) basis set, or PWs with a plane-wave cutoff of 340 eV.  Gas phase structures for the full Chl~\emph{a} and \emph{b} monomers (see Figure~\ref{MonomerStructure}(a) \emph{full}$\cdot$\textbf{1}$^+$) and the Chl~\emph{a} dimer (see Figure~\ref{DimerStructure}) were based on those of Refs.~\citenum{Bruce1} and \citenum{Chla2}, respectively.  To determine the influence of the hydrocarbon chain and the tetramethylammonium charge tag (N(CH$_3$)$_4^+$ or \textbf{1}$^+$) on the optical absorbance of Chl~\emph{a} and \emph{b} monomers we have also considered structures with the hydrocarbon chain cut from C$_{20}$H$_{39}$ to C$_{5}$H$_{9}$ with and without the N(CH$_{3}$)$_{4}^+$ charge tag, as shown in Figure~\ref{MonomerStructure} (a) \emph{cut}$\cdot$\textbf{1}$^+$ and (b) \emph{cut}, respectively.

Each structure was optimized using the ASE code \cite{ASE0,ASE} until a maximum force less than 0.03~eV/\AA{} was obtained. All structures were modelled in supercells with more than 6~\AA{} of vacuum using non-periodic boundary conditions, that is, the electron density and wave functions were set to zero at the cell boundary. This prevents long-range interactions between repeated images that would affect a periodic calculation.  More importantly, the use of non-periodic boundary conditions has been previously shown to be essential for modelling charged structures in gas phase \cite{CH4Gold}.

Optical absorbance spectra were modelled using the imaginary part of the dielectric function, $\Im[\varepsilon(\omega)]$, from LCAO-TDDFT calculations in the frequency domain (\LCAOTDDFTkomega{}) \cite{GlanzmannTDDFTRPA,PreciadoSWCNTs,LCAOTDDFTKeenan,LCAOTDDFTkomega}, and using the Fourier transform of the dipole moment, $\mathcal{F}\{d_m(t)\}$, from LCAO-TDDFT in the real time domain (\LCAOTDDFTrt{}) \cite{LCAOTDDFT}.

The optical absorbance from \LCAOTDDFTkomega{} in the $\eq$ direction, neglecting local field effects, is given by \cite{PreciadoSWCNTs,LCAOTDDFTKeenan}
\begin{equation}
\Im[\varepsilon(\omega)] = \frac{1}{\Omega} \sum_{nm} \frac{4\pi\eta[f(\varepsilon_{m}) - f(\varepsilon_{n})]}{(\omega - \varepsilon_{n} + \varepsilon_{m} - \Deltax)^2 + \eta^2}
 \left|\frac{\eq\cdot\langle\psi_{n}|{\mathbf{\nabla}} |\psi_{m}\rangle}{\varepsilon_{n} - \varepsilon_{m}+\Deltax}\right|^2\label{Imeps}
\end{equation}
where $\Omega$ is the supercell volume, $\eta$ is the electronic broadening,  $f$ is the Fermi--Dirac function,  $\varepsilon_{n}$ is the eigenenergy and $\psi_{n}$ is the KS wave function of the $n$th orbital, and $\Deltax$ is the derivative discontinuity correction to the eigenenergies \cite{GLLBSC} from the exchange part of the GLLB-SC functional given by \cite{PreciadoSWCNTs}
\begin{equation}
    \Deltax = \frac{8\sqrt{2}}{3\pi^2} \sum_{n=1}^N \left(\sqrt{\varepsilon_{N+1} - \varepsilon_n} - \sqrt{\varepsilon_{N} - \varepsilon_n}\right)\langle\psi_{N+1}|\frac{\psi_{n}^*\psi_{n}^{}}{\rho}|\psi_{N+1}\rangle,\label{eq:Deltax}
\end{equation}
where $N$ is the number of electrons.  Incorporating this derivative discontinuity correction has been shown to provide better agreement with experimental band gaps \cite{Castelli2012GLLBsc2} and optical absorption spectra \cite{PreciadoSWCNTs,LCAOTDDFTKeenan}.

The matrix elements in Eq.~\eqref{Imeps} may be expressed as \cite{GlanzmannTDDFTRPA}
\begin{equation}
  \langle\psi_{n}|{\mathbf{\nabla}} |\psi_{m}\rangle\!=\!\sum_{\mu\nu}\!c_{\nu n}^*c_{\mu m}^{} \langle \tilde{\phi}_{\nu}|\Proj^\dagger{\mathbf{\nabla}}\Proj|\tilde{\phi}_{\mu}\rangle,\label{matrixelements}
\end{equation}
 where the sum is over $\mu \equiv \{i,a\}$ for the $i$th state centered on atom $a$, $\tilde{\phi}_{\mu}$ are the localized basis functions, so that $|\tilde{\psi}_{m}\rangle = \sum_{\mu}c_{\mu m}|\tilde{\phi}_{\mu}\rangle$ with coefficients $c_{\mu m}$ for the $m$th KS wave function,
and $\Proj$ is the PAW transformation operator \cite{PAW,PAW2,CarstenThesis}
\begin{equation}
  \Proj = 1 + \sum_{a i} \left(|\varphi_{i}^a\rangle - |\tilde{\varphi}_{i}^a\rangle\right)\langle \tilde{p}_i^a|,
\end{equation}
where $\tilde{\varphi}_i^a$ and $\varphi_i^a$ are the pseudo and all-electron partial waves for state $i$ on atom $a$ within the PAW formalism, and $|\tilde{p}_{i}^a\rangle$ are the smooth PAW projector functions. 

Since the matrix elements of Eq.~\eqref{matrixelements} must already be calculated to obtain the forces at the DFT level, the calculation of $\Im[\varepsilon(\omega)]$ using Eq.~\eqref{Imeps} simply involves the multiplication of previously calculated matrices.  For this reason, the calculation of $\Im[\varepsilon(\omega)]$ using \LCAOTDDFTkomega{} is very efficient, with scaling of $\BigO(NM^2)$ or better \cite{slug}, where $N$ is the number of KS wavefunctions and $M \geq N$ is the total number of basis functions used in the LCAO calculation.  This provides a significant speed-up compared to the $\BigO(N^5)$ scaling of the Casida \LCAOTDDFTromega{} formalism \cite{LCAOTDDFT, Casida1995,TDDFTRevCasida2009}.  We found ten unoccupied KS wave functions per chlorophyll molecule were already sufficient to converge the spectra up to 3.5~eV.  This is because local field effects may be safely neglected in the optical limit for gas phase structures \cite{DuncanGrapheneTDDFTRPA,GlanzmannTDDFTRPA}.

\begin{figure*}
  \centering
  \includegraphics[width=1.5\columnwidth]{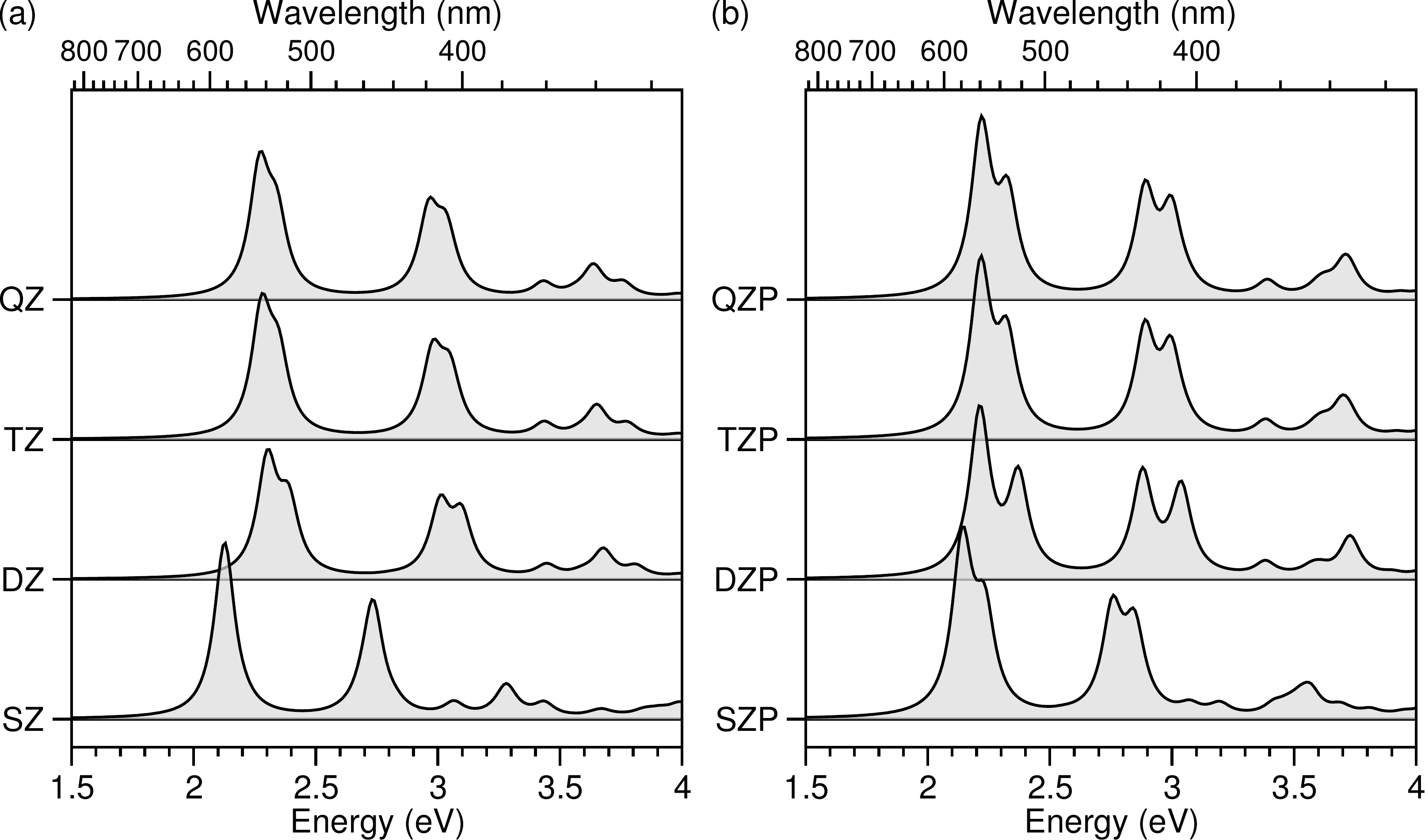}
  \caption[Basis Set Dependence of the Optical Absorption Spectra for Chl \textit{a}]{Dependence of the \LCAOTDDFTkomega{} optical absorption spectra of Chl \textit{a} on the single-$\zeta$ (SZ), double-$\zeta$ (DZ), triple-$\zeta$ (TZ) and quadruple-$\zeta$ (QZ) LCAO basis sets (a) without and (b) with polarization (P) functions.}
  \label{fig:basis_set_dependence}
\end{figure*}

The optical absorbance from \LCAOTDDFTrt{} is given by the Fourier transform of the dipole moment, $d_m$, so that
\begin{equation}
  \FT\{d_m(t)\} = \frac{1}{\pi \kick}\int \sin(\omega t) e^{- \frac{t^2}{2\sigma^2}} [d_m(t) - d_m(0)] dt
\end{equation}
where $\kick = 1\times 10^{-5}$ is the strength of the ``kick'' in the $\eq$ direction, and $\sigma$ is the peak broadening.  The dipole moment is then obtained from the time dependent charge density $\rho(\rv,t)$ using
\begin{equation}
  d_m(t) = -\iiint \rho(\rv,t) (\rv \cdot \eq) dV. 
\end{equation}
The time dependent charge density is simply $\rho(\rv,t) = - e \sum_n f(\varepsilon_n) |\psi_n(\rv,t)|^2$, where $\psi_n(\rv,t)$ are the time-dependent KS wave functions which are obtained by propagating
\begin{equation}
  i\frac{\partial}{\partial t} \psi_n(\rv,t) = \hamKS(t)\psi_n(\rv,t),\label{TDDFT}
\end{equation}
where $\hamKS(t) = -\frac{1}{2}\nabla^2 + v_{\mathrm{KS}}[\rho(\rv,t)](\rv,t)$ is the KS Hamiltonian for independent electrons, and $v_{\mathrm{KS}}$ is the KS potential functional.

Employing LCAO to represent the time-dependent KS wave functions, we may express Eq.~\eqref{TDDFT} as \cite{LCAOTDDFT}
\begin{equation}
 \sum_{\mu} i \langle \tilde{\phi}_{\nu} |\Proj^\dagger \Proj |\tilde{\phi}_{\mu}\rangle \frac{\partial}{\partial t} c_{\mu n}(t) = \sum_{\mu,m}\langle \tilde{\phi}_{\nu}| \Proj^\dagger \hamKS^{nm}(t) \Proj^\dagger | \tilde{\phi}_{\mu}\rangle c_{\mu m}(t).\label{LCAOTDDFT}
\end{equation}
After applying the initial kick, the time propagation is performed using the reliable and numerically stable semi-implicit Crank--Nicolson method \cite{CrankNicolson}, as described in Ref.~\citenum{LCAOTDDFT}.

As was the case for \LCAOTDDFTkomega{}, each step of the \LCAOTDDFTrt{} is very efficient and involves solving the system of linear equations, Eq.~\eqref{LCAOTDDFT}, with ScaLAPACK \cite{slug}.  For this reason, the calculation of $\FT\{d_m(t)\}$ using \LCAOTDDFTrt{} scales as $\BigO(NM^2)$.  Although this is worse than the $\BigO(N G)$ scaling of \RSTDDFTrt{}, where $G$ is the number of grid points, the constant prefactor for the grid propagation is so large that even for systems with thousands of electrons it is outperformed by several orders of magnitude by \LCAOTDDFTrt{} \cite{LCAOTDDFT}.

All spectra have been calculated with intrinsic widths of 50 meV.  In the frequency domain we employed an electronic broadening of $\eta = 50$~meV to the individual Lorentzian peaks, while in the real time domain we convoluted the dipole moment with a Gaussian of width  $\sigma \approx 13$~fs.

The \LCAOTDDFTrt{} spectra remained stable with a 0.01~fs time step, which was sufficient to resolve the shape of the dipole spectra.  Such a time step is an order of magnitude greater than the 0.002~fs time step required for real space time propagation calculations to maintain stability \cite{Joaquim}.  We were thus easily able to propagate up to 80~fs, twice that recorded with real space methods \cite{Joaquim}.

To provide insight into the spatial distribution of the exciton, we employ the electron hole density difference, $\Delta \rho(\rv, \omega) = \rho_h(\rv, \omega) + \rho_e(\rv,\omega)$, as described in Refs.~\citenum{Catechol,PreciadoSWCNTs,LCAOTDDFTKeenan}.  Here, the electron/hole densities $\rho_{e/h}(\rv)$ are obtained by averaging the two-particle exciton density $\rho_{ex}(\rv_e,\rv_h)$, with respect to the hole/electron coordinate $\rv_{h/e}$.  We calculate the electron hole density difference $\Delta \rho (\rv, \omega)$ for the first four bright excitations using the transitions from \LCAOTDDFTkomega{}.  Since these molecular transitions are both well separated and composed of single transitions $m \to n$ at the \LCAOTDDFTkomega{} level, we find $\Delta \rho(\rv, \omega_{m n}) \approx e(|\psi_n|^2 - |\psi_m|^2)$, where $\psi_{m}$ and $\psi_{n}$ are the KS orbitals corresponding to the hole and electron, respectively, and $\omega_{m n} \approx \varepsilon_n - \varepsilon_m + \Deltax$ is the energy of the transition $m \to n$, including the derivative discontinuity correction $\Deltax$ of Eq.~\eqref{eq:Deltax}.

\section{RESULTS AND DISCUSSION}\label{Sect:ResultsDiscussion}

In Figure \ref{fig:basis_set_dependence} we show the optical absorption spectrum of the neutral \textit{cut} structure of Chl \textit{a} calculated using either multiple-$\zeta$ basis sets (SZ, DZ, TZ and QZ) or the polarized versions of the basis sets (SZP, DZP, TZP and QZP) within the LCAO mode of the \textsc{gpaw} code. We systematically increase the number of functions to assess the sensitivity of the optical absorbance and observe that the spectrum converges differently for basis sets with and without polarization functions.

It can be seen in Figure \ref{fig:basis_set_dependence}(a) that the SZ basis set yields a red-shifted spectrum compared to the other spectra, the TZ and QZ basis sets spectra are almost the same, and the DZ basis set yields the same energies of the Q and Soret bands compared to those of the TZ and QZ but the peaks within the band are more separated than in the TZ or QZ spectra. The polarized multiple-$\zeta$ basis sets are shown in Figure \ref{fig:basis_set_dependence}(b) and we observe that the SZP basis set also yields a red-shifted spectra compared to other spectra and there are more peaks above $3$~eV. The spectra obtained using TZP and QZP are nearly identical and the DZP basis set yields the same energy of the Q and the Soret bands although the peaks within the bands are a little bit separated when compared to TZP and QZP.

Even though it seems that the spectrum of Chl \emph{a} will not change if we increase the number of functions of the basis set, we cannot rely solely on these results to claim that an LCAO basis set will guarantee an accurate description of the optical absorption spectra. This is because LCAO basis sets cannot be systematically converged to the complete basis set limit, for molecules of this size. This is clear from comparing the result of the inclusion of polarization functions with the LCAO basis sets in Figure~\ref{fig:basis_set_dependence}. Although double-$\zeta$ basis sets are converged with respect to the number of radial functions, the inclusion of polarization systematically alters the spectra. For this reason, we must compare the spectra of Chl \emph{a} and Chl \emph{b} obtained using LCAO and a PW representations of the KS wavefunctions.

\begin{figure}
  \centering
  \includegraphics[width=0.8\columnwidth]{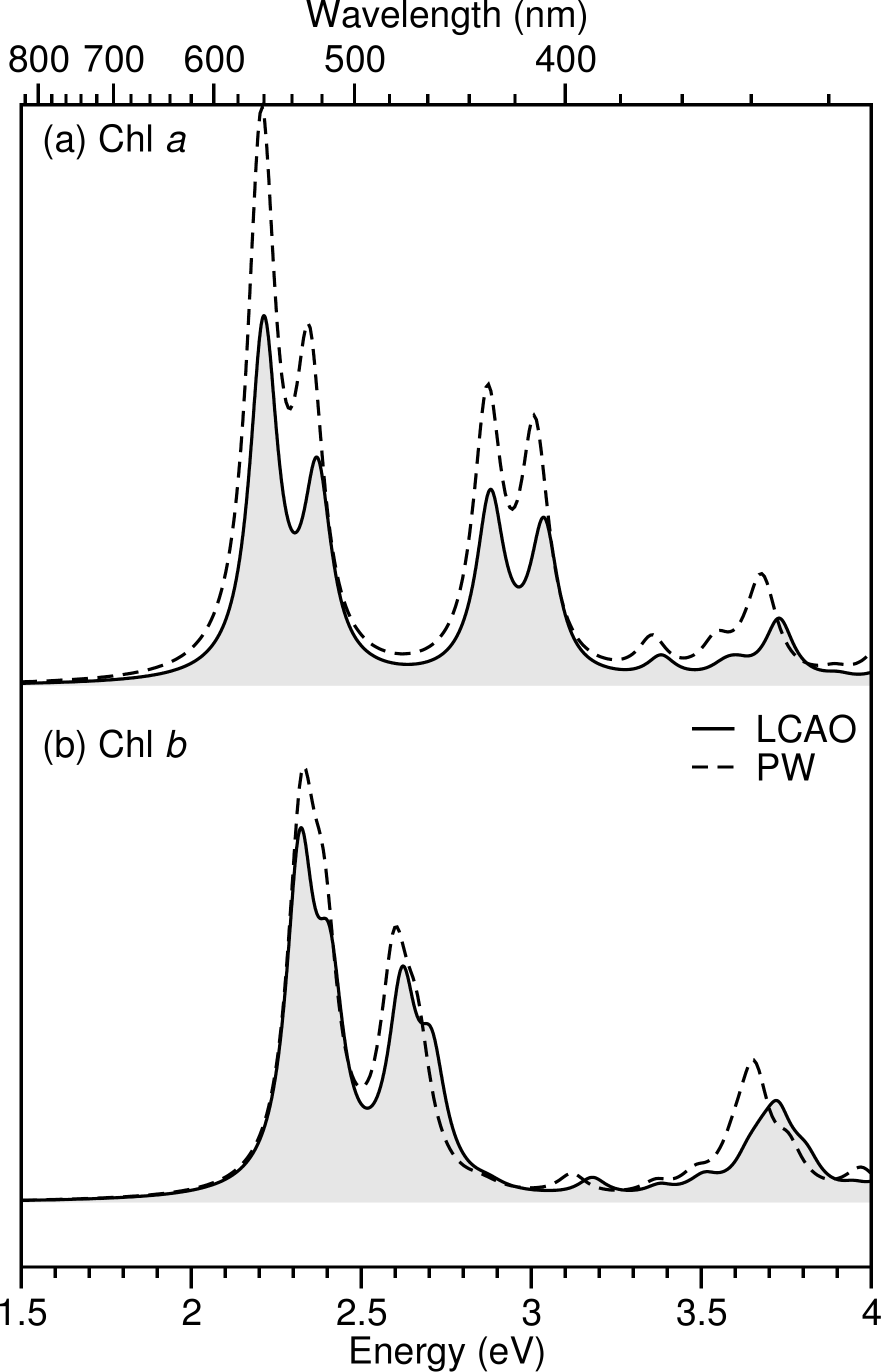}
  \caption[LCAO and PW Optical Absorption Spectra for Chl \emph{a} and \emph{b}]{Optical absorption of Chl \textit{a} and \emph{b} employing LCAO (solid lines, \LCAOTDDFTkomega{}) and PW (dashed lines, \PWTDDFTkomega{}) representations of the Kohn-Sham orbitals.}
  \label{fig:lcaopw}
\end{figure}

In Figure \ref{fig:lcaopw}, we show the optical absorption spectra of the \textit{cut} structures of both Chl \textit{a} and Chl \textit{b} calculated using either the DZP basis set within the LCAO mode, or using a PW representation of the KS wavefunctions that ensures converged optical absorption spectra. We find that the DZP basis set yields an optical absorption spectrum with transition energies very close to those of the spectrum calculated using PW. We observe that the LCAO spectra is only slightly (much smaller than the $0.1$ eV DFT accuracy) red-shifted compared to the PW spectra, but underestimates the intensities of the transitions. Nevertheless, we can affirm that the atomic basis set with double-$\zeta$ and polarization functions, DZP, is sufficient to ensure semi-quantitative agreement of the optical absorption spectra for the Chl \textit{a} and Chl \textit{b} with the PW calculations.

The DZP basis sets have been shown to be sufficient (and often necessary) to converge to the results calculated with PW representations of the KS wavefunctions in other works as well \cite{BenchmarkPaper, GPAWLCAO}. From hereon we shall restrict consideration to the DZP basis set and simply refer to our calculations as LCAO.

\begin{figure}
  \includegraphics[width=\columnwidth]{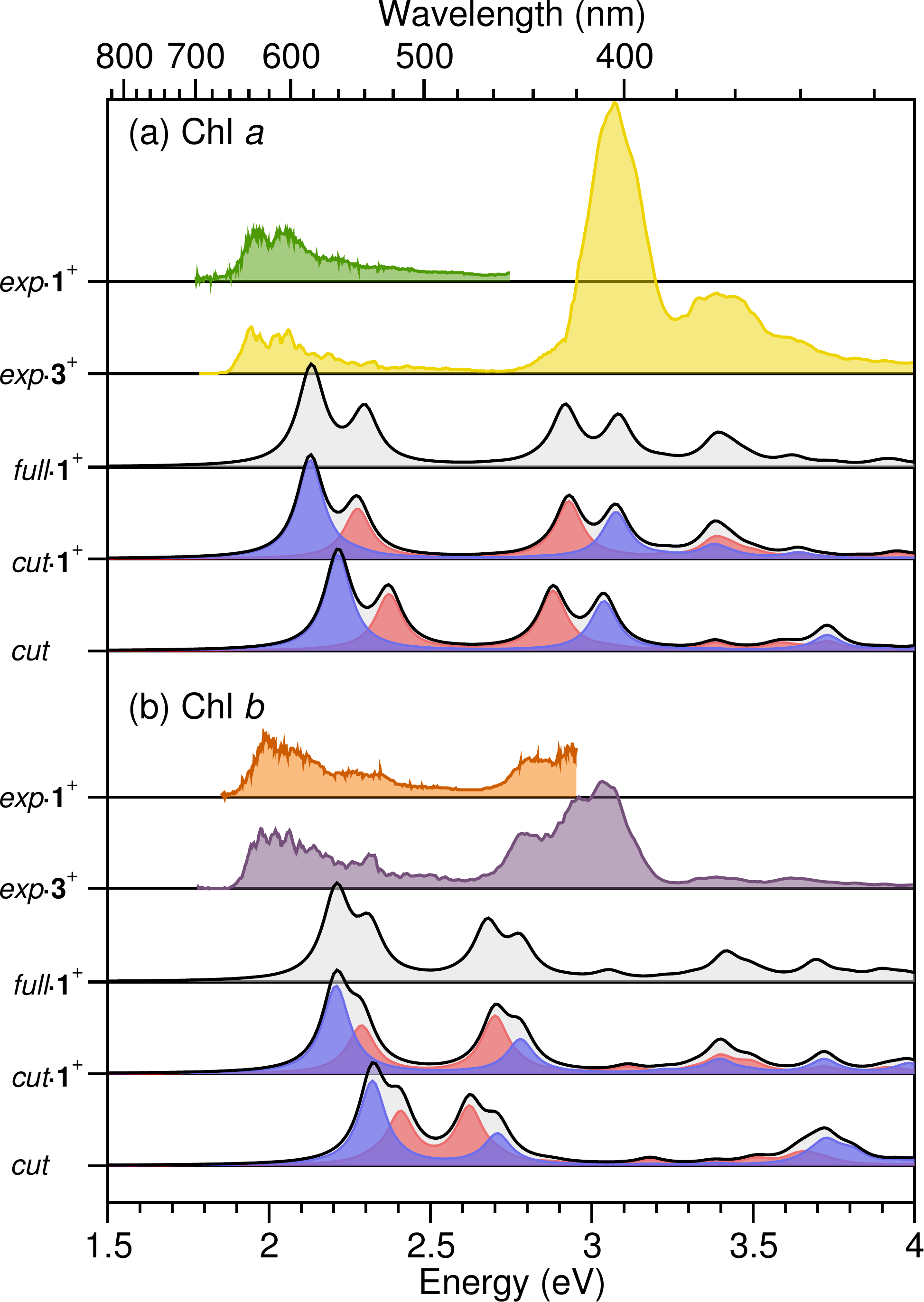}
  \caption{Optical absorbance spectra, $\Im[\varepsilon(\omega)]$, from \LCAOTDDFTkomega{} for (a) Chl~\emph{a} and (b) Chl~\emph{b} \emph{full}$\cdot$\textbf{1}$^+$, \emph{cut}$\cdot$\textbf{1}$^+$,  and \emph{cut} structures, with spectra decomposed into absorption in the $x$ (blue) or $y$ (red) polarization directions \cite{Chlorophylls,Moss88}, as shown in Figure~\ref{MonomerStructure}. These are compared with those measured experimentally with tetramethylammonium and acetylcholine charge tags, \emph{exp}$\cdot$\textbf{1}$^+$ and \emph{exp}$\cdot$\textbf{3}$^+$, respectively, from Refs.~\citenum{Bruce1} and \citenum{Bruce2}.
  }\label{Fig2TDDFTRPA}
\end{figure}

\begin{figure*}
  \includegraphics[width=\textwidth]{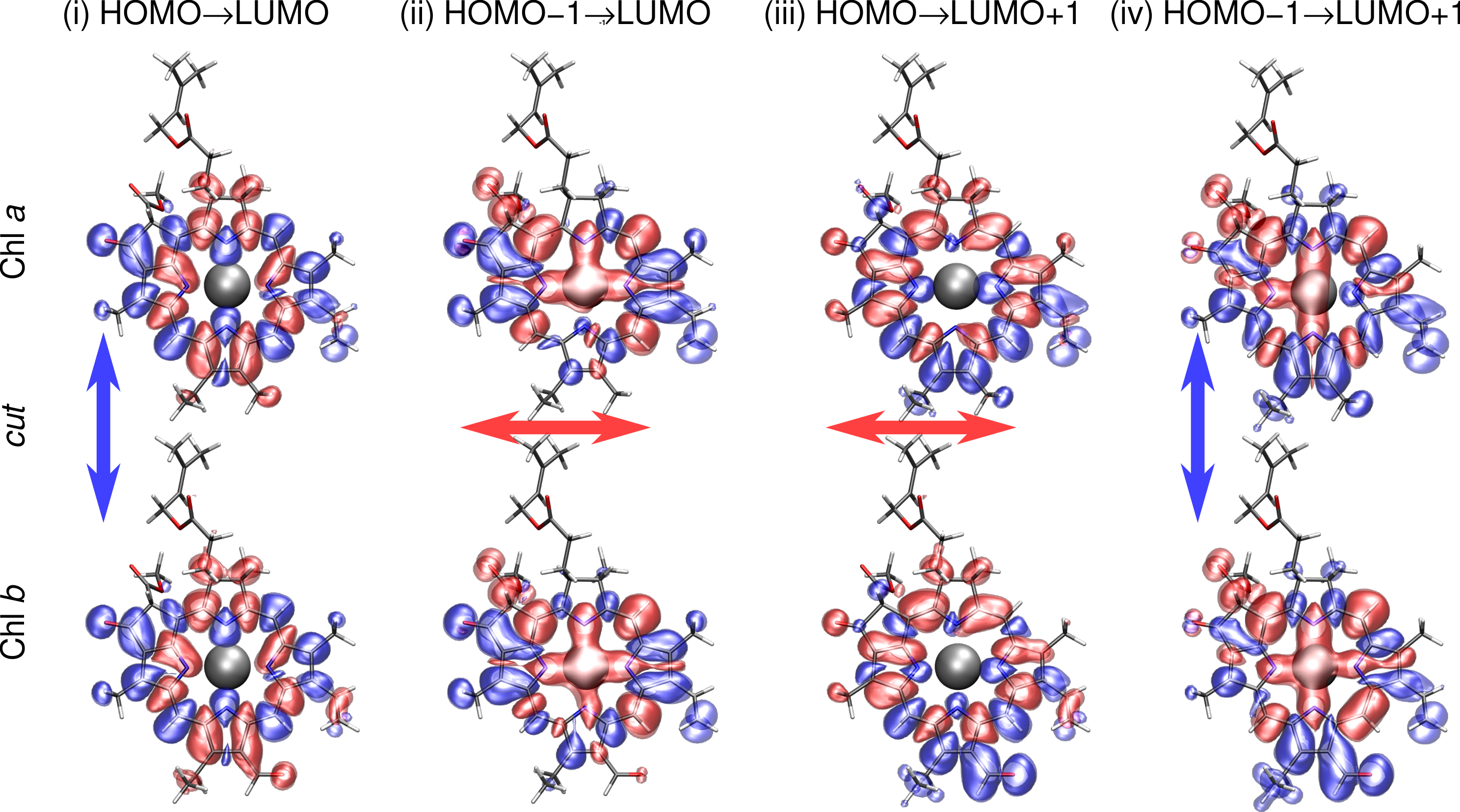}
  \caption{Isosurfaces of the electron (blue) and hole (red) density difference $\Delta\rho(\rv, \omega_i)= \pm 0.001~e/$\AA$^3$ of the first four excitations at $\omega_i$ of neutral Chl~\emph{a} and \emph{b} with a cut hydrocarbon chain (\emph{cut}) (i) HOMO $\rightarrow$ LUMO, (ii) HOMO$-1$ $\rightarrow$ LUMO, (iii) HOMO $\rightarrow$ LUMO$+1$, and (iv) HOMO$-1$ $\rightarrow$ LUMO$+1$.  Absorption is along the N--Mg--N bonds either in the $x$ (blue arrows) or $y$ (red arrows) polarization directions \cite{Chlorophylls,Moss88}.  Mg, C, O, N, and H atoms are depicted in silver, grey, red, blue, and white.}\label{ExcitationsNeutral}
\end{figure*}

\begin{figure*}[!t]
  \includegraphics[width=\textwidth]{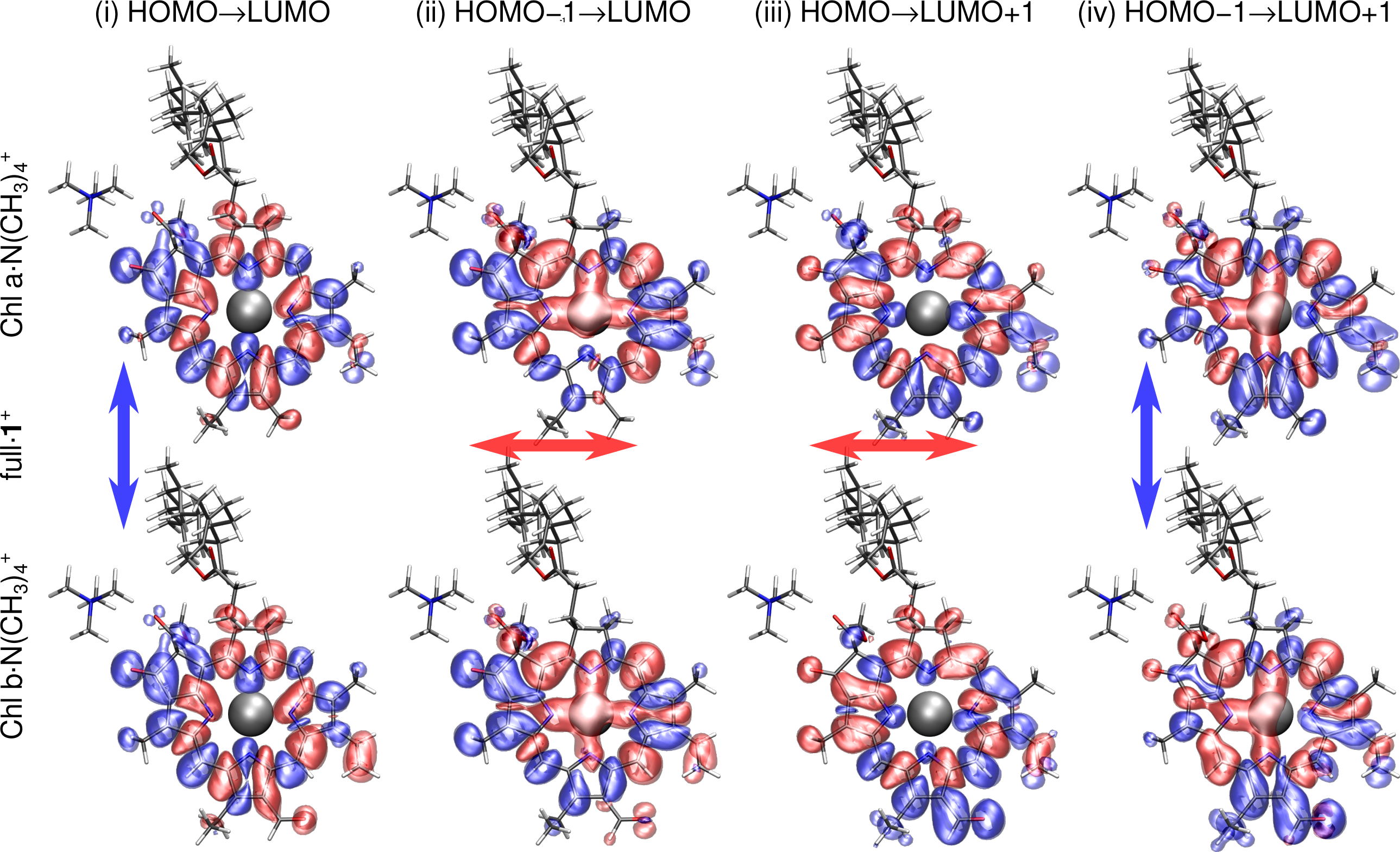}
  \includegraphics[width=\textwidth]{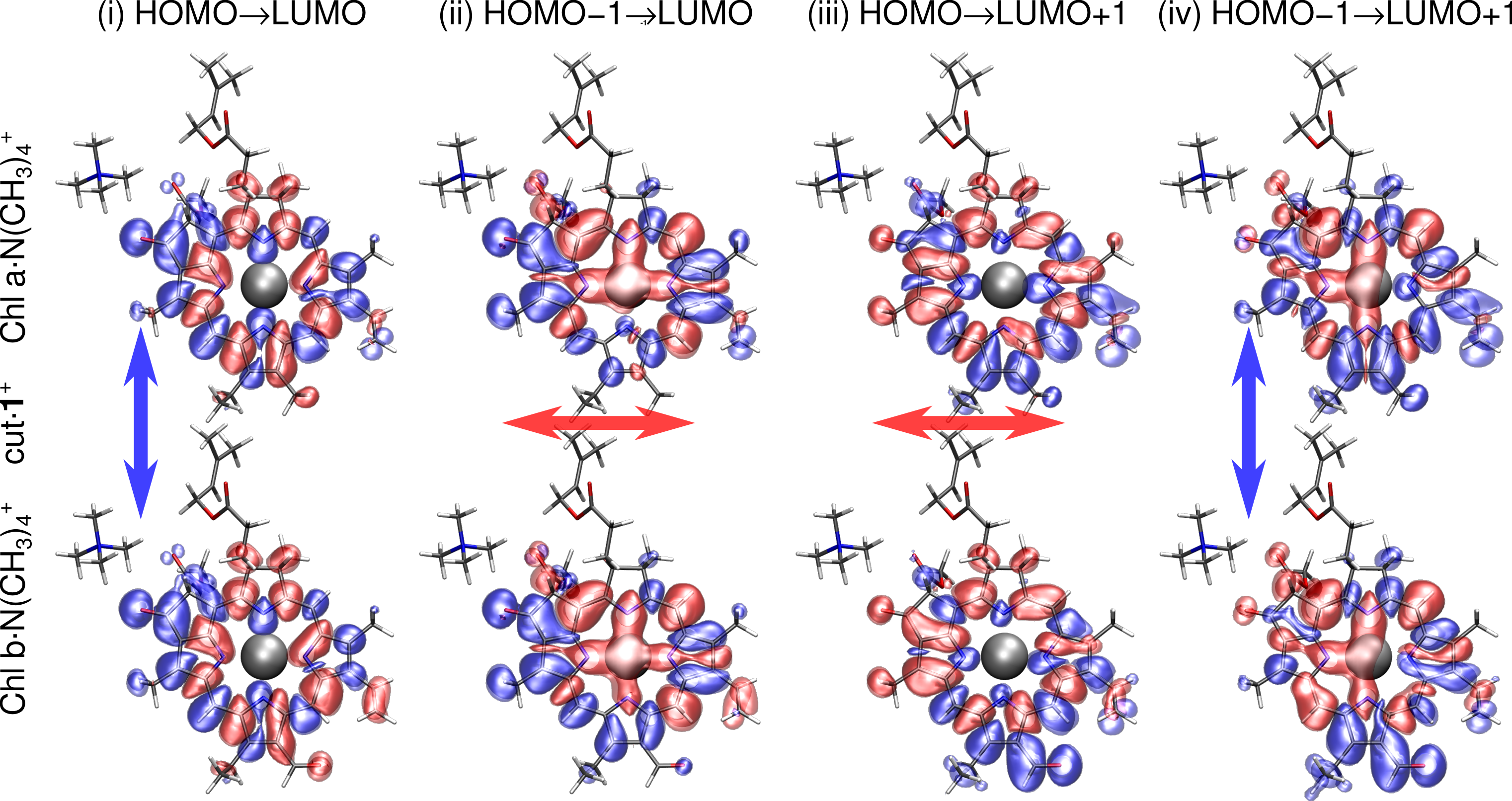}
  \caption{Isosurfaces of the electron (blue) and hole (red) density difference $\Delta\rho(\rv,\omega_i) = \pm 0.001 e/$\AA$^3$ of the first four excitations $\omega_i$ of charge tagged Chl~\emph{a}$\cdot$N(CH$_3$)$_4^+$ and Chl~\emph{b}$\cdot$N(CH$_3$)$_4^+$ with a full (full$\cdot$\textbf{1}$^+$, upper panel) and cut (cut$\cdot$\textbf{1}$^+$, lower panel) hydrocarbon chain (i) HOMO $\rightarrow$ LUMO, (ii) HOMO$-1$ $\rightarrow$ LUMO, (iii) HOMO $\rightarrow$ LUMO$+1$, and (iv) HOMO$-1$ $\rightarrow$ LUMO$+1$.  Mg, C, O, N, and H atoms are depicted in silver, grey, red, blue, and white.}\label{ExcitationsCharged}
  \end{figure*}

In Figure \ref{Fig2TDDFTRPA} we plot the spectra obtained from \LCAOTDDFTkomega{} including the derivative discontinuity correction $\Deltax$ for the Chl~\emph{a} and \emph{b} monomers shown in Figure~\ref{MonomerStructure}.   The onset of the calculated spectra are in semi-quantitative agreement with the experimental spectra obtained with either the monocationic tetramethylammonium \textbf{1}$^+$ or acetylcholine \textbf{3}$^+$ tag. The difference between the maximum of the Q band and the calculated first excitation energy is less than $0.2$ eV and the difference between the maximum of the Soret band and the calculated fourth excitation is less than $0.1$ eV. Also, the spectra of the \textit{cut}$\cdot$\textbf{1}$^+$ structure shows the same qualitative behavior as the \textit{full}$\cdot$\textbf{1}$^+$ structure, suggesting that none to negligible changes in the optical absorption spectra are caused by the carbon chain. However, when the charge tag is removed, that is, regarding the \textit{cut} structure, the Q band peaks are blue shifted and the band gap is widened, whereas the Soret band peaks are red shifted.

For Chl \textit{b}, Figure \ref{Fig2TDDFTRPA}(b), the calculated spectra for the \emph{full}$\cdot$\textbf{1}$^+$ structure is also in semi-quantitative agreement with the experimental spectra.  Specifically, the peaks of the Q band and the Soret band of the \textit{full}$\cdot$\textbf{1}$^+$ structure are blue and red shifted, respectively, when compared to the experimental data, with a difference between the Q band maximum and the calculated first excitation of $0.23$ eV and a difference between the Soret band maximum and the calculated fourth excitation of $0.26$ eV. Again, the spectrum of the \textit{cut}$\cdot$\textbf{1}$^+$ structure is qualitatively the same as the spectrum of the \textit{full}$\cdot$\textbf{1}$^+$, reinforcing the idea that the carbon chain has no impact in the optical absorbance and that it should be centered on the Mg atom and the chlorin ring. When removing the charge tag (\textit{cut} structure) the Q band peaks are again blue shifted, and the band gap is widened, whereas the Soret band peaks are red shifted.

The intensities of the excitations of the Soret band are being underestimated about $75\%$ for Chl \textit{a} and $52\%$ for Chl \textit{b}, comparing the \textit{full}$\cdot$\textbf{1}$^+$ to the experimental data. Also, the intensity of the Soret band peaks is less compared to that of the Q band peaks. This is not the case for the experimental data. Although the relative intensities are adjusted, that is, the Soret band peaks intensities are increased by applying the GLLB-SC correction to the spectra following Eq.~\ref{Imeps}, they are still underestimated.

Such a difference can be explained by the fact that we are neglecting charge transfer excitations at the linear density response level. The higher intensity of these peaks is often attributed in the literature to this type of charge transfer excitation \cite{Bruce1}, which we are unable to model at the \LCAOTDDFTkomega{} level. In this way, our underestimation of the Soret band intensity provides indirect insight into the nature of the experimentally observed peaks.

In Figure~\ref{ExcitationsNeutral} we plot the spatial distribution of the electron and hole densities for the first four bright transitions of the Chl~\emph{a} and \emph{b} \emph{cut} structures.   As shown in Figure~\ref{Fig2TDDFTRPA}, the first and fourth excitations are induced by optical absorption along the N--Mg--N bond in the $x$ polarization direction \cite{Moss88,Chlorophylls} (blue arrows in Figure~\ref{ExcitationsNeutral}), while the second and third excitations are induced by optical absorption along the $y$ polarization direction \cite{Moss88,Chlorophylls} (red arrows in Figure~\ref{ExcitationsNeutral}), as expected.  In each case, the excitations are $\pi\rightarrow\pi$ transitions involving the two highest occupied molecular orbitals (HOMOs) and the two lowest unoccupied molecular orbitals (LUMOs).
 
 For both species the electron density of the first and third excitations has weight on the $2p_z$ levels of the N atoms parallel to the direction of excitation towards the Mg atom.  Conversely, the hole density of the second and third excitations on both Chl~\emph{a} and \emph{b} has significant weight on the $2p_z$ levels of both the N and Mg atoms. Overall, the first and third transitions tend to move charge density from the chlorin ring to either its edges or towards the central Mg atom, whereas the second and fourth transitions tend to move charge from the central Mg atom to the edge of the chlorin ring along the direction of excitation.  

 Although the excitations for Chl~\emph{a} and \emph{b} are qualitatively the same, we do notice differences in the region of the methyl/aldehyde.  Specifically, the electron of the third and fourth transitions has significantly more weight around the aldehyde group of Chl~\emph{b} compared to the methyl group of Chl~\emph{a}.  Also, the hole density of the second and fourth transitions has weight on all four N atoms for Chl~\emph{b}, whereas the electron density has some weight on one of the N atoms perpendicular to the direction of excitation.

 Finally, the excitations of the charge tagged species with a cut (\emph{cut}$\cdot$\textbf{1}$^+$) and full (\emph{full}$\cdot$\textbf{1}$^+$)  hydrocarbon chain, shown in Figure~\ref{ExcitationsCharged}, are basically the same as those for the neutral species (\emph{cut} shown in Figure~\ref{ExcitationsNeutral}).  This provides additional evidence justifying the experimental use of charge tagged Chl~\emph{a} and \emph{b} molecules to describe the optical absorption of the neutral isolated species \cite{Bruce1,Bruce2}.

\begin{figure}
    \includegraphics[width=\columnwidth]{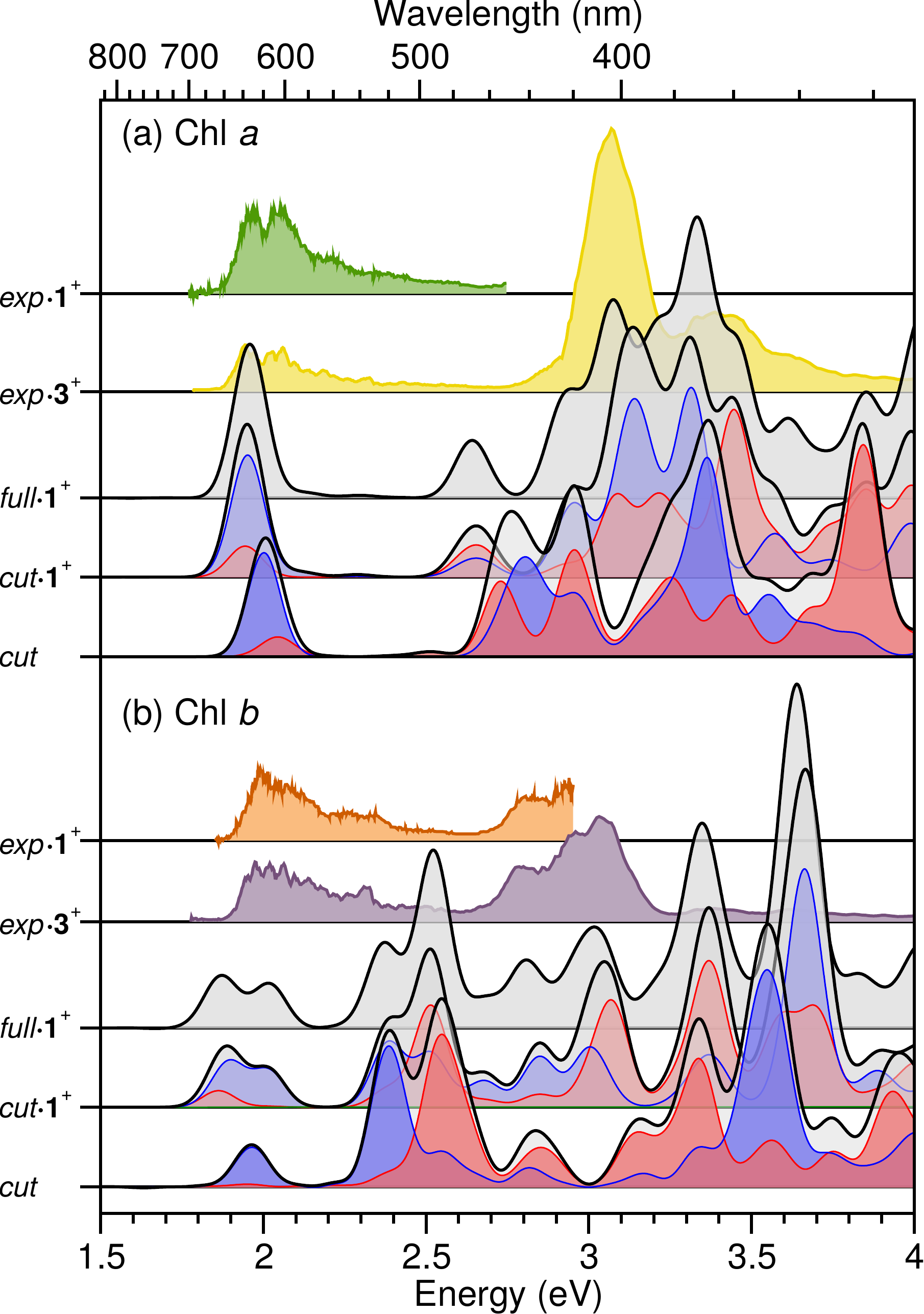}
  \caption{Optical absorbance spectra, $\FT\{d_m(t)\}$, from \LCAOTDDFTrt{} for (a) Chl~\emph{a} and (b) Chl~\emph{b} \emph{full}$\cdot$\textbf{1}$^+$, \emph{cut}$\cdot$\textbf{1}$^+$, and \emph{cut} structures, with spectra decomposed into absorption in the $x$ (blue) or $y$ (red) polarization directions \cite{Moss88,Chlorophylls}, as shown in Figure~\ref{MonomerStructure}.  These are compared with those measured experimentally with tetramethylammonium and acetylcholine charge tags, \emph{exp}$\cdot$\textbf{1}$^+$ and \emph{exp}$\cdot$\textbf{3}$^+$, respectively, from Refs.~\citenum{Bruce1} and \citenum{Bruce2}.  }
  \label{Fig3TDDFTprop}
\end{figure}

In Figure~\ref{Fig3TDDFTprop} we directly compare the spectra obtained from \LCAOTDDFTrt{} for the Chl~\emph{a} and \emph{b} monomers shown in Figure~\ref{MonomerStructure} with those measured experimentally from Refs.~\citenum{Bruce1} and \citenum{Bruce2}.  The Q-band energy or onset of the \LCAOTDDFTrt{} spectra agrees quantitatively with that measured experimentally for both Chl~\emph{a} and \emph{b} in Refs.~\citenum{Bruce1} and \citenum{Bruce2}.  However, whereas we found distinct Q band transitions with polarizations in the $x$ and $y$ directions for \LCAOTDDFTkomega{} (see Figure~\ref{Fig2TDDFTRPA}), with \LCAOTDDFTrt{} we find the Q band transitions coincide in energy.  Moreover, for Chl~\emph{b} \emph{cut}$\cdot$\textbf{1}$^+$, we find that whereas there are two Q-band peaks, they both have polarization in the $x$ direction, from ring \textbf{B} to ring \textbf{D} along the N--Mg--N bond.  This splitting of the Q band peak in the $x$ direction and the weak Q band peak in the $y$ direction are both suppressed once the charge tag is removed (see Figure~\ref{Fig3TDDFTprop}(b) \emph{cut}).

While we find removing the charge tag induces a blue shift of the Q band for Chl~\emph{a} and \emph{b}, as was the case for \LCAOTDDFTkomega{}, we find the Soret band is less affected by the charge tag with \LCAOTDDFTrt{}.  In the region of the Soret band \LCAOTDDFTrt{} yields many peaks from 2.5--4~eV.  This may be related to the difficulty in converging higher energy peaks with TDDFT-$t$, although our time propagation of $\sim 80$~fs should be sufficient to converge these peaks.

Overall, we find the spectra from \LCAOTDDFTrt{} provide excellent agreement with the experimentally measured spectra with tetramethylammonium (\emph{exp}$\cdot$\textbf{1}$^+$) and acetylcholine (\emph{exp}$\cdot$\textbf{3}$^+$) charge tags.  This provides further evidence that the exact structure of the charge tag has little to no impact on the onset of the spectra \cite{Bruce1,Bruce2}.  Moreover, the spectra exhibit rather minor changes upon removing the charge tag, providing further justification for using charge tagged Chl~\emph{a} and \emph{b} to describe the optical absorption of the neutral isolated species \cite{Bruce1,Bruce2}.

\begin{figure}
  \includegraphics[width=\columnwidth]{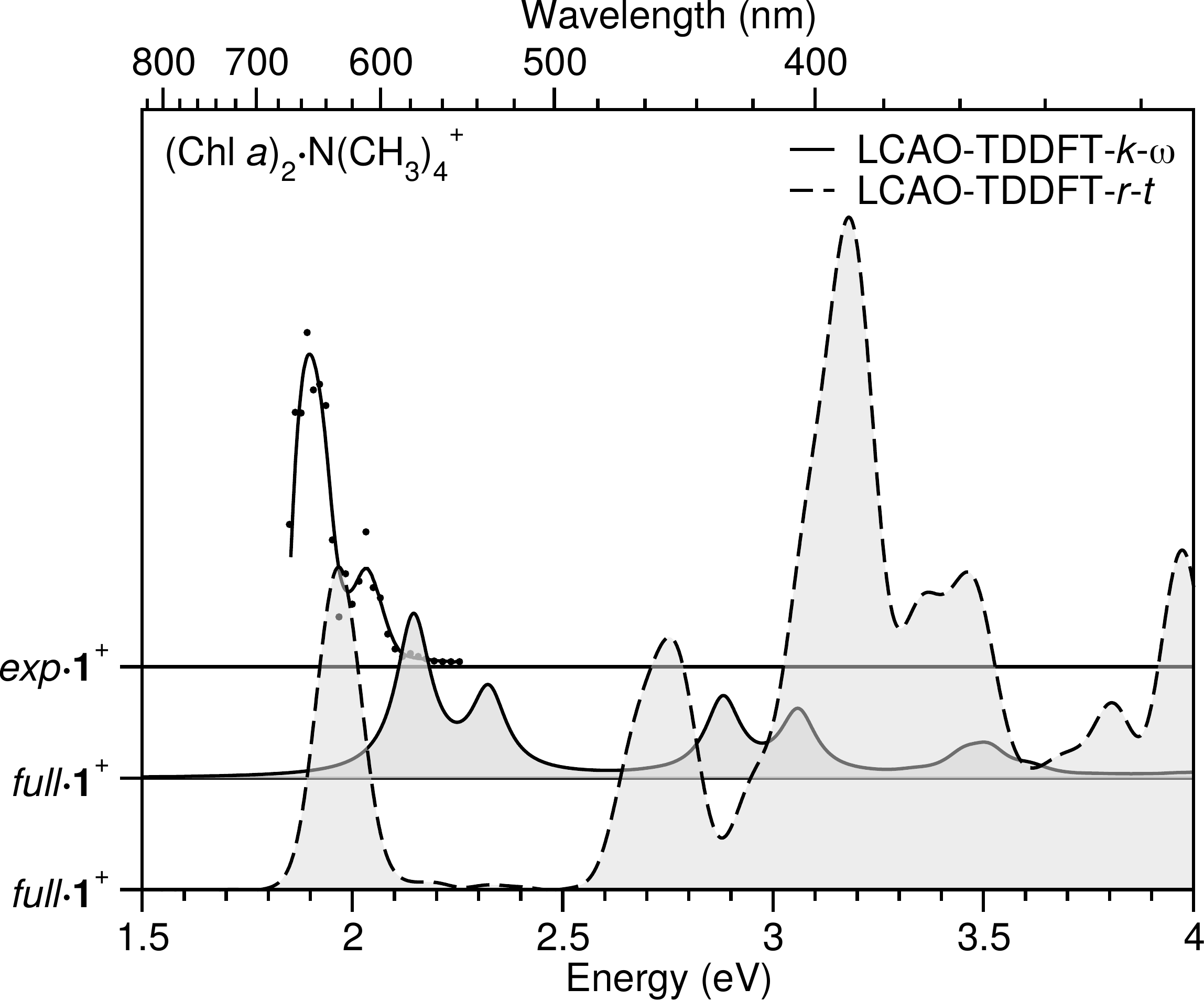}
\caption{Optical absorbance spectra from \LCAOTDDFTkomega{} (solid lines,  $\Im[\varepsilon(\omega)]$), and \LCAOTDDFTrt{} (dashed lines,  $\mathcal{F}\{d_m(t)\}$) for the (Chl~\emph{a})$_2\cdot$N(CH$_3$)$_4^+$ structure (\emph{full}$\cdot$\textbf{1}$^+$) shown in Figure~\ref{DimerStructure} and as measured experimentally (\emph{exp}$\cdot$\textbf{1}$^+$,$\bullet$) from Ref.~\citenum{Chla2}.  
}\label{Dimer:fgr}
\end{figure}

Going beyond the monomer, in Figure~\ref{Dimer:fgr} we directly compare the optical absorbance spectra for the Chl~\emph{a} dimer measured experimentally from Ref.~\citenum{Chla2}, obtained from \LCAOTDDFTkomega{}, $\Im[\varepsilon(\omega)]$, and \LCAOTDDFTrt{}, $\mathcal{F}\{d_m(t)\}$.  In both cases, the experimental onset of the peak is reproduced semi-quantitatively, with the derivative discontinuity corrected \LCAOTDDFTkomega{} results somewhat blue shifted by about $0.2$~eV.   However, as was the case for the monomer, the experimentally observed splitting of the Q band is clearly visible with \LCAOTDDFTkomega{} but absent from the \LCAOTDDFTrt{} Chl~\emph{a} spectrum.

Overall, these results show that increasing the system size, as we go from the monomer to the dimer, \LCAOTDDFTkomega{} and \LCAOTDDFTrt{} provide the same level of accuracy.  For \LCAOTDDFTkomega{}, this means that after applying the derivative discontinuity correction $\Deltax \approx 0.5$~eV, the spectra agree semi-quantitatively with the measured spectra for both the Chl~\emph{a} monomer and dimer.   For \LCAOTDDFTrt{}, this correction is not required to obtain a quantitative description of the experimental spectra for both the Chl~\emph{a} monomer and dimer. Altogether, this suggests that for even larger systems, such as the LHC II, we may reasonably expect both \LCAOTDDFTkomega{} and \LCAOTDDFTrt{} to continue to perform well.  

\begin{table}
\caption{Q band energies in eV and wavelengths in nm for Chl~\emph{a} and \emph{b} monomers and dimers with and without the \textbf{1}$^+$ charge tag.}\label{Table1}
\begin{tabular}{@{}ll@{}l@{\hspace{1em}}l@{}c@{}l@{\hspace{1em}}l@{}}
  \hline\hline
\multicolumn{1}{l@{ }}{method} & 
\multicolumn{1}{l@{}}{structure} & 
\multicolumn{1}{c}{(eV)}&
\multicolumn{1}{c}{(nm)}&
\multicolumn{1}{c}{}&
\multicolumn{1}{c}{(eV)}&
\multicolumn{1}{c}{(nm)}\\\hline
&&
  \multicolumn{2}{c}{Chl~\emph{a}}&&
  \multicolumn{2}{c}{Chl~\emph{b}}\\\cline{3-4}\cline{6-7}
  \\[-0.75em]
  PID$^a$ & \emph{exp}$\cdot$\textbf{1}$^+$ & 1.97 & 629 &&
  1.99 & 623\\
  \multirow{2}{*}{$\frac{4\pi\omega}{c} \Im[\alpha(\omega)]^b$} & \emph{full}$\cdot$\textbf{1}$^+$ & 2.08 & 596 &&
  2.15 & 577 \\
  & \emph{full} & 2.04 & 608 &&
  2.13 & 582 \\[0.5em]
  & \emph{full}$\cdot$\textbf{1}$^+$ & 1.96& 633 &&
  1.87, 2.02 & 662, 615 \\
  $\FT\{d_m(t)\}^c$ & \emph{cut}$\cdot$\textbf{1}$^+$ & 1.95 & 636&&
  1.89, 2.00 & 657, 621\\
  &\emph{cut} & 2.01 & 618&&
  1.96 & 631\\[0.5em]
  & \emph{full}$\cdot$\textbf{1}$^+$ & 2.13, 2.30  & 582, 540 &&  2.21, 2.31 & 562, 537\\
  $\Im[\varepsilon(\omega)]^d$ & \emph{cut}$\cdot$\textbf{1}$^+$ & 2.13, 2.27 & 583, 545 &&  2.21, 2.29 & 562, 542\\
  &\emph{cut} & 2.21, 2.37 & 561, 524 &&  2.32, 2.41 & 534, 515\\
    $\Im[\varepsilon(\omega)]^e$ & \emph{cut} & 2.21, 2.35 & 561, 528 &&  2.33, 2.39 & 533, 518\\[0.5em]
&&
\multicolumn{2}{c}{(Chl~\emph{a})$_2$}\\\cline{3-4}
\\[-0.75em]
PID$^f$ & \emph{exp}$\cdot$\textbf{1}$^+$ &  1.90\hspace{0.5em} & 652 \\
$\frac{4\pi\omega}{c}\Im[\alpha(\omega)]^g$ & \emph{full}$\cdot$\textbf{1}$^+$ & 2.05 & 605 \\
$\FT\{d_m(t)\}^c$ & \emph{full}$\cdot$\textbf{1}$^+$ & 1.97 & 630\\
$\Im[\varepsilon(\omega)]^d$ & \emph{full}$\cdot$\textbf{1}$^+$ & 2.15, 2.32 & 578, 534\\\hline\hline
\multicolumn{7}{p{\columnwidth}}{\footnotesize$^a$Photoinduced dissociation from Ref.~\citenum{Bruce1}\nocite{Bruce1}. $^b$TD-CAM-B3LYP/Def2-SVP from Ref.~\citenum{Bruce1}. $^c$\LCAOTDDFTrt{} from this work. $^d$\LCAOTDDFTkomega{} from this work. $^e$\PWTDDFTkomega{} from this work. $^f$Photoinduced dissociation from Ref.~\citenum{Chla2}\nocite{Chla2}.  $^g$TD-CAM-B3LYP/Def2-SVP from Ref.~\citenum{Chla2}.}\\
\end{tabular}
\end{table}

In Table~\ref{Table1} we provide a direct comparison of the Q-band energy (first and second transitions shown in Figures~\ref{ExcitationsNeutral} and \ref{ExcitationsCharged}) for Chl~\emph{a}, Chl~\emph{b}, and (Chl~\emph{a})$_2$ obtained from PID measurements, TD-CAM-B3LYP/Def2-SVP calculations of the Casida \LCAOTDDFTromega{} type \cite{Casida1995,TDDFTRevCasida2009} using the imaginary part of the dynamic polarizability $\alpha(\omega)$ ($\frac{4\pi \omega}{c}\Im[\alpha(\omega)]$), \LCAOTDDFTrt{} ($\FT\{d_m(t)\}$), \PWTDDFTkomega{} and \LCAOTDDFTkomega{} ($\Im[\varepsilon(\omega)]$).  Overall, we find excellent agreement between PID, \LCAOTDDFTrt{}, and  TD-CAM-B3LYP/Def2-SVP.  In fact, our \LCAOTDDFTrt{} calculations perform somewhat better than  TD-CAM-B3LYP/Def2-SVP for describing the Q band energies of Chl~\emph{a}, Chl~\emph{b}, and (Chl~\emph{a})$_2$.  These LCAO-TDDFT results, along with its computational speed-up, provide strong motivation for the future use of both \LCAOTDDFTkomega{} and  \LCAOTDDFTrt{} for describing larger systems such as the LHC~II.

\section{CONCLUSIONS}\label{Sect:Conclusions}

 We have shown that \LCAOTDDFTkomega{} \cite{GlanzmannTDDFTRPA,PreciadoSWCNTs,LCAOTDDFTKeenan} and \LCAOTDDFTrt{} \cite{LCAOTDDFT} calculations are both significantly faster than their \PWTDDFTkomega{},  TD-CAM-B3LYP/Def2-SVP, and \RSTDDFTrt{} counterparts and also provide a similar level of accuracy to alternative TDDFT methods in the description of the optical absorption spectrum of the key photosynthetic light-harvesting molecules Chl~\emph{a} and \emph{b} and the dimeric system (Chl~\emph{a})$_2$.  Specifically, our \LCAOTDDFTkomega{} and \LCAOTDDFTrt{} calculations provide both a semi-quantitative and qualitative description of the spectra of chlorophyll monomers and dimers, with the former yielding a significant reduction in memory requirements compared to \PWTDDFTkomega{} and the latter yielding an order of magnitude speed-up compared to \RSTDDFTrt{} calculations due to the increased numerical stability of \LCAOTDDFTrt{}.  These results pave the way for future studies applying these highly efficient and accurate LCAO-TDDFT methods to the characterization of biomacromolecules, such as the LHC~II, and the full understanding of the physical and chemical processes involved in photosynthesis.


 \acknowledgments
 
This work employed the Imbabura cluster of Yachay Tech University, which was purchased under contract No.\ 2017-024 (SIE-UITEY-007-2017), and the Quinde I supercomputer of Public Company Yachay E.~P., which was implemented under contract No.~0051-2015, corresponding to Component No.~7 of Group No.~2, Re-YACHAY-018-2015.  D.J.M.\ thanks the management and operation team of the Quinde I Supercomputer for their help and assistance in this project.  
A.H.L.\ acknowledges funding from the European Union's Horizon 2020
research and innovation program under grant agreement no.~676580 with
The Novel Materials Discovery (NOMAD) Laboratory, a European Center of
Excellence; the European Research Council (ERC-2010-AdG-267374);
Spanish grant (FIS2013-46159-C3-1-P); and Grupos Consolidados
(IT578-13).
The Coimbra Chemistry Centre is supported by FCT, through the Project PEst-OE/QUI/UI0313/2014 and POCI-01-0145-FEDER-007630. B.F.M.\ thanks the Portuguese Foundation for Science and Technology (projects CENTRO-01-0145-FEDER-000014 and POCI-01-0145-FEDER-032229) and the Donostia International Physics Centre for financial support.

\end{document}